\documentclass[pra,twocolumn,a4paper,superscriptaddress]{revtex4}

 \usepackage{latexsym}
 \usepackage{amsmath}
 \usepackage{amsfonts}
 \usepackage{graphicx}
 \usepackage{amssymb}
 \usepackage{subfigure}
 \usepackage{color}
 \usepackage{mathtools}

 \newcommand{\ket}[1]{\ensuremath{|#1\rangle}}
 \newcommand{\bra}[1]{\ensuremath{\langle #1 |}}

 \newcommand{\bc}{\begin{center}}
 \newcommand{\ec}{\end{center}}

 \newcommand{\ii}{i}
 \newcommand{\DP}{\Delta_p}
 \newcommand{\DC}{\Delta_c}

 \newcommand{\bl}{\beta_{L}}

 \newcommand{\bs}{\beta_{T}}

 \usepackage[utf8]{inputenc}

\begin{document}

\title{Controlled light shaping via phase dependent electromagnetically induced transparency}

\author{Sandeep \surname{Sharma}}
\email{sandeep.sharma@iitg.ernet.in}
\affiliation{Department of Physics, Indian Institute of Technology Guwahati,
Guwahati- 781039, Assam, India}

\author{Tarak N. \surname{Dey}}
\email{tarak.dey@iitg.ernet.in}
\affiliation{Department of Physics, Indian Institute of Technology Guwahati,
Guwahati- 781039, Assam, India}
\date{\today}

\begin{abstract}
We explore optical manipulation of sculpted light based on phase dependent electromagnetically induced transparency through a five level atomic system.
A transverse magnetic field (TMF) and a suitable spatially inhomogeneous control field can be used to create a spatial probe transparency modulation at a desired location.
Such transparency modulation is the principle behind the shaping of the light.
Further the beam propagation equation shows that the control field induced selective phase information can be imprinted on the probe beam.
Hence this controlled light shaping paves a new  way for optical tweezers, high contrast imaging and micromachining.
\end{abstract}
\maketitle
\begin{figure}[h]
\includegraphics[width=8cm,height=8cm,keepaspectratio]{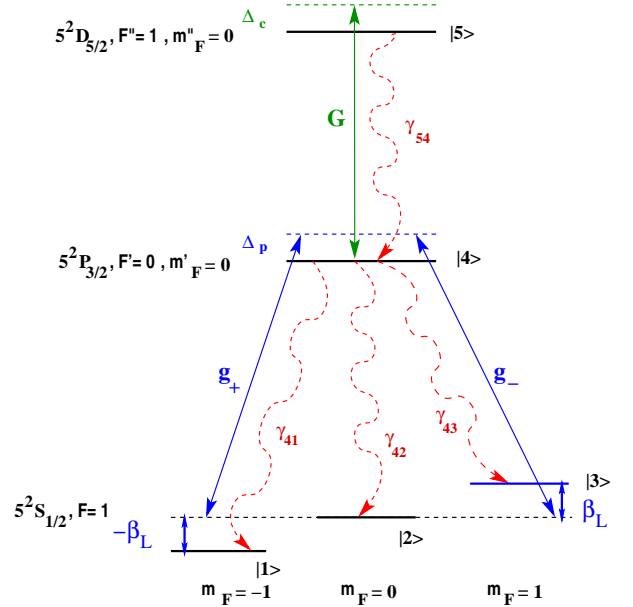}
\caption{\label{fig:Fig1} (Color online) 
Schematic diagram of the inverted-Y level system of $^{87}Rb$ atoms. A strong control field with Rabi frequency $G$ couples the transition $\ket{4}\leftrightarrow\ket{5}$. The transitions $\ket{4}\leftrightarrow\ket{1}$ and $\ket{4}\leftrightarrow\ket{3}$ are coupled by the probe field components with Rabi frequency $g_{_{+}}$ and $g_{_{-}}$, respectively}.
\end{figure}

Spatial manipulation of beams carrying orbital angular momentum (OAM) have recently gained a lot of attention owing to their immense application in microtrapping and alignment \cite{Ashkin,Gahegan,Paterson}, optical micromanipulation \cite{Woerdemann}, optical communications \cite{Gibson}, and biosciences \cite{Stevenson}. 
Control of the beam structure is generally carried out using various optical elements such as axially symmetric polarization element \cite{Sakamoto}, porro-prism \cite{Litvin}, spatial light modulator \cite{Jesacher,Moreno, Arnold}, etc. 
The fundamental behind such structural modulation of the beams rely on the superposition principle in which two beams with equal but opposite OAM interfere to create a beam with structured spatial patterns. 
On the contrary, Radwell {\it et al.} have introduced a very interesting technique for creating structured beams using quantum coherence effects in atomic medium \cite{Radwell}. 
In the experiment, they considered a cold rubidium system in closed-loop Hanle configuration \cite{Renzoni}, driven by a structured probe \cite{Marrucci} and TMF. 
The presence of TMF along with a phase structured probe induces phase dependent atomic coherences (PDACs) \cite{Sharma1}. 
These PDACs initiate a periodic oscillation of the absorption in the azimuthal plane, which is a key factor in the creation of structured beams. 
Manipulation of oscillating absorption in the transverse plane using various spatially dependent control fields can play a major role in the generation of structured beams with different spatial forms.  \\
In this paper, we propose a new scheme for selectively generating various structured beams by exploiting OAM of probe beam and spatially dependent control field.
An inverted Y-type five level atomic system consisting of two excited states $\ket{5}$ and $\ket{4}$ and three metastable states $\ket{1}$, $\ket{2}$, and $\ket{3}$ as shown in Fig.(\ref{fig:Fig1}) is deliberately introduced for the realisation of such a scheme.
A linearly polarized strong control field is tunned to the atomic transition $\ket{4}\leftrightarrow\ket{5}$, whereas two orthogonal components of the probe field are coupled to the remaining dipole allowed transitions $\ket{4}\leftrightarrow\ket{3}$ and $\ket{4}\leftrightarrow\ket{1}$.
The electric fields for both probe(p) and control(c) beams propagating along the $z$-axis, are described as
\begin{align}
\vec{E}(\vec{r},t)= \hat{x}\sum\limits_{j=p,c} \mathcal{E}_{j}(\vec{r})~e^{- i\left(\omega_j t-  k_j z\right)} + {c.c.}\,,
\end{align}
where $\mathcal{E}_{j}(\vec{r})$ , $\omega_j$, and $k_j$  are the slowly varying envelope, the frequency and the wave vector of the fields, respectively.
Phase dependent opical properties can be activated by employing an arbitrary magnetic field $\vec{B}= B(\cos\theta~{\bf \hat{z}}+\sin\theta~{\bf \hat{x}})$ 
on the Zeeman sublevels which form a closed loop system.
The transverse component of magnetic field (TMF) $B\sin\theta$ plays a key role in the creation of phase dependent atomic coherence.
However, the longitudinal component $B\cos\theta$ of the magnetic field gives rise to splitting between the Zeeman states $\ket{1}$ and $\ket{3}$.
Further, we work in the weak TMF regime {\it i.e.,} $\theta<<\pi/2$, which ensures that the probe polarization remains unaffected and the quantization axis remains aligned along the $z$-axis \cite{Margalit}. 
In this regime, the interaction Hamiltonian can be written explicitly for the system under the dipole and rotating wave approximation, respectively, as
\begin{align}
\label{interaction_hamiltonian}
H_{I}=&\hbar\Delta_p\ket{4}\bra{4} + \hbar(\Delta_c+\Delta_p)\ket{5}\bra{5} \nonumber  \\
&-\hbar\left(g_{+}\ket{4}\bra{1} + g_{-}\ket{4}\bra{3} + G\ket{5}\bra{4}\right ) \nonumber \\
&+\hbar\beta_{L}(\ket{3}\bra{3}-\ket{1}\bra{1})+ \hbar\beta_{T}(\ket{1}\bra{2}+\ket{2}\bra{3})+\text{H.c.},
\end{align}
where $g_{\pm}=(\vec{d}_{\pm}\cdot\vec{\cal E}_{\pm})e^{ik_pz}/\hbar $ are the Rabi frequencies of the right (left) circular polarized component of the probe and ${\cal E}_{\pm}= \mathcal{E}_{p}/\sqrt{2}$.
The control Rabi frequency is denoted as $G=(\vec{d}_{c}\cdot\vec{\cal E}_{c})e^{ik_cz}/\hbar $.
The term $\beta_{T}=\beta_{0}\sin\theta/\sqrt{2}$ describes the coupling strength between \ket{2} to \ket{1} and \ket{2} to \ket{3}.
The magnitude of Zeeman shift between the ground levels \ket{1} and \ket{3} is given by $\beta_{L}=\beta_{0}\cos\theta$.
The magnetic parameter is $\beta_0=g_{F}\mu_{B}B/\hbar$, where $\mu_B$ and $g_{F}$ are the Bohr magneton and Lande g-factor, respectively.
The probe and the control fields detunings from the respective resonant transitions can be defined as $\Delta_p=\omega_p-\omega_{42}$ and $\Delta_c=\omega_c-\omega_{54}$ .
 In the presence of various coherent and incoherent processes,  the density matrix formalism is used for the analysis of the atomic population and coherences.
 Therefore,  the corresponding Liouville equation is given as
\begin{align}
\label{master}
\dot{\rho}=-\frac{\ii}{\hbar}\left[H_{I},\rho\right]+\mathcal{L}_{r}\rho+\mathcal{L}_{c}\rho\,.
\end{align}
The last two terms in Eq.(\ref{master}) describe all incoherent processes which are given as
\begin{align}
\label{decay2}
\mathcal{L}_{r}\rho = &-\sum\limits_{i=1}^3 \frac{\gamma_{4i}}{2}\left(\ket{4}\bra{4}\rho-2\ket{i}\bra{i}\rho_{44}+\rho\ket{4}\bra{4}\right)\nonumber \\
&- \frac{\gamma_{54}}{2}\left(\ket{5}\bra{5}\rho-2\ket{4}\bra{4}\rho_{55}+\rho\ket{5}\bra{5}\right)\,, \nonumber \\
\mathcal{L}_{c}\rho = &-\sum\limits_{j=1}^3\sum\limits_{j\ne i=1}^3 \frac{\gamma_{c}}{2}\left(\ket{j}\bra{j}\rho-2\ket{i}\bra{i}\rho_{jj}+\rho\ket{j}\bra{j}\right)\,\nonumber,
\end{align}
where the decay rates are expressed as $\gamma_{41}=\gamma_{42}=\gamma_{43}=\gamma/3$, and $\gamma_{54}=0.1\gamma$. Here $\gamma$ is the spontaneous decay rate of state $\ket{4}$. 
The collisional dephasing rate of the metastable ground states is given by $\gamma_c$.

We now evaluate an analytical expression for the medium susceptibility $\chi_{_{\pm}}$ in the steady state limit.
The linear susceptibilities of the probe field components can be expressed as 
\begin{equation}
\label{chi_41-43}
\chi_{_{\pm}}=\frac{\mathcal{N}|d_{\pm}|^2}{{\hbar}\gamma}\left(\frac{N_{\pm}}{D}+\beta^{2}_{T}\frac{N_{0}}{D}\frac{g_{_{\mp}}}{g_{_{\pm}}}\right)\,,
\end{equation}
where
\begin{align}
N_{\pm}=&\mp 2\Gamma_{4i}\beta^{3}_{L}{|G|}^{2}\rho^{(0)}_{ii} \pm 2\Gamma_{4i}\beta_{L}\beta^{2}_{T}{|G|}^{2}\rho^{(0)}_{22}\nonumber \\
&+2\ii \Gamma_{4i} \Gamma_{54}({|G|}^{2}+\Gamma_{4i}\Gamma_{54})(\beta^{2}_{L}-\beta^{2}_{T})\rho^{(0)}_{ii}\,, \nonumber \\
N_{0}=&-2 \ii \Gamma_{4i} \Gamma_{54}({|G|}^{2}+\Gamma_{4i}\Gamma_{54})(\rho^{(0)}_{ii}-\rho^{(0)}_{22})\,, \nonumber \\
D=&(-2{|G|}^{2}(\beta^{2}_{L}+2\beta^{2}_{T})+({|G|}^{2}+\Gamma_{4i}\Gamma_{54})^{2})(\beta^{2}_{L}-\beta^{2}_{T})\,. \nonumber
\end{align}
The label $i$ is 1 for index $+$ and 3 for index $-$. The atomic density of the medium is given by $\mathcal{N}$.
The population of Zeeman sublevels  are expressed as $\rho^{(0)}_{11}=\rho^{(0)}_{33}= (\bs/\bl)^{2}$, and $\rho^{(0)}_{22}= 1-2(\bs/\bl)^{2}$.
Eqs.(\ref{chi_41-43}) clearly show that the coupling of both probe components are present due to $\beta_T$, which is necessary for observing the phase dependent susceptibility.
Further, the phase dependent characteristics of the medium can be well understood by considering the spatial variations in the probe field components {\it i.e.} $g_{\pm}(\vec{r})= g(r,z)e^{\pm(il\phi)}$.
The transverse profile of the probe beam is $g(r,z)$, and $l$, $\phi$ are the OAM and phase carried by probe beam, respectively. 
The medium susceptibilities in the presence of an inhomogeneous probe beam can be defined as
\begin{align}
\chi_{_{\pm}}= \frac{\mathcal{N}|d_{\pm}|^2}{{\hbar}\gamma}\left(\frac{N_{\pm}}{D}+\beta^{2}_{T}e^{\mp 2il\phi}\frac{N_{0}}{D}\right)\label{chi_41}\,. 
\end{align}
The above analytical expressions show the dependence of medium susceptibilities on TMF strength $\beta_T$, OAM $l$, transverse phase $\phi$ and control field intensity. All these parameters play a decisive role in the creation and manipulation of the phase dependent medium response.

\begin{figure}[h]
\centering\includegraphics[width=8.5cm,height=7.5cm]{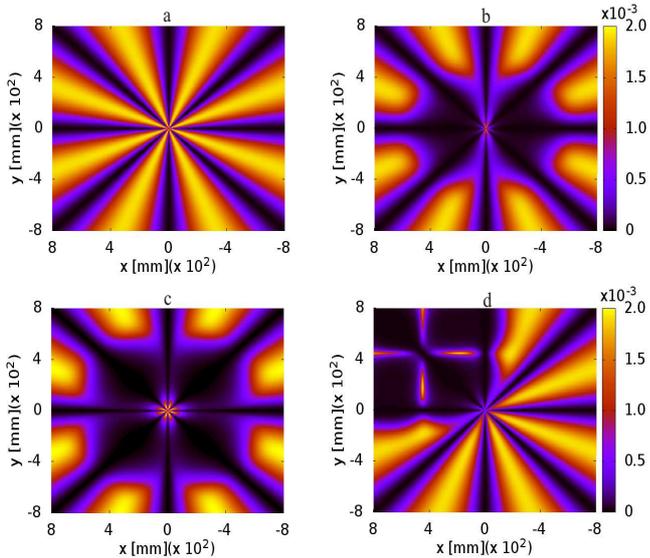}
\caption{ \label{fig:Fig2} (Color online) 
Spatial modulation of absorption of the $\hat{\sigma}_{+}$ polarisation component is plotted against the two orthogonal axes $x$ and $y$. 
Panel (a) display the behaviour of absorption in absence of control field. 
Panel (b), (c) and (d) exhibit the phase dependent absorption in presence of $HG_{10}$, $HG_{11}$ and shifted $HG_{11}$ control beam. 
The parameters of control beam are $G_{0}=1.0\gamma$ and $w_{c}=30~\mu$m. 
Other parameters are chosen as ${\mathcal{N}}=2~\times~10^{12} atoms/cm^{3}$, $\Gamma_{41}=0.5{\gamma}$, $\Gamma_{54}=0.5{\gamma}$, $\DP =0$, $\DC =0$, $\beta_{0}=0.01{\gamma}$, $\gamma_{c}=10^{-7}{\gamma}$, $\theta=\pi/12$, $p=0$, $q=0$, and OAM of probe beam $l=4$, except for panel (d) where $p=45~\mu$m, and $q=45~\mu$m.}
\end{figure}

We first study the phase-dependent response of the medium in absence of control field by evaluating Eq.~(\ref{chi_41}).
Fig.(\ref{fig:Fig2})(a) shows that the absorption of $\hat{\sigma}_{+}$ component of the probe varies periodically along the $x-y$ plane in the presence of both azimuthal probe phase and TMF. 
These periodic variations of absorption in the azimuthal plane initiate a spatial transparency window at a specific angular position $n\pi/l$.
Hence the exponential factor $e^{\mp 2il\phi}$ in Eq.~(\ref{chi_41}) gives rise to spatial inhomogeneity of the medium. 
The number of transparency windows formed in the azimuthal plane can be determined by the $2l$ factor in the exponential term. 
So for OAM $l=4$, 8 transparency windows are created in the transverse plane as shown in Fig.(\ref{fig:Fig2})(a).
A similar $2l-$fold symmetric absorption pattern is found for the $\hat{\sigma}_{-}$ component of the probe at resonance $\DP =0$, as explicitly shown by Eq.~(\ref{chi_41}).
Such phase dependent dynamics of the probe absorption is the essence of structured beam generation.
However, we observe in Fig.(\ref{fig:Fig2})(a) that a huge probe absorption poses as an obstacle for the formation of structured beams. 
A drastic change in the character of spatial inhomogeneity of the medium from symmetric to asymmetric can be made possible by introducing a spatially dependent control beam.
This asymmetric transparency window paves the way for the optical manipulation of the spatial mode at a desired location.
Basically the inverted $Y$-level system can be decomposed as two $\equiv$ type systems in the absence of TMF.
The $\equiv$ type systems are well known to display EIT line shape \cite{Banacloche} at two photon resonance condition $\DP = \DC =0$.
Therefore the transparency induced by the control beam as well as the TMF holds the key to the formation of asymmetric probe absorption pattern in the medium.
Before discussing the effect of spatial inhomogeneity of the control field on probe susceptibility, we examine how control field intensity reduces the resonant probe absorption.
In this case, the spatial variation of probe absorption is same as in Fig.(\ref{fig:Fig2})(a) except that the absorption is reduced by a factor of $10^{-2}$.
Hence, the control field renders  an otherwise  opaque medium highly transparent for the probe beam.
For accomplishment of spatial absorption modulation, we have chosen the following transverse profile of the control field:
\begin{align}
G(x,y,z)=& G_{0}\left(\frac{w_{c}}{w_{c}(z)}\right) H_{m}\left(\frac{(x-p)\sqrt{2}}{w_{c}(z)}\right)H_{n}\left(\frac{(y-q)\sqrt{2}}{w_{c}(z)}\right)\nonumber \\ 
& e^{-((x-p)^{2}+(y-q)^{2})/w_{c}^{2}(z)}e^{ik((x-p)^{2}+(y-q)^{2})/2R_{c}(z)}\nonumber \\ 
& e^{i(kz-(m+n+1)\tan^{-1}(z/z_{0}))}\,.
\end{align}
where $H_{i}$ is Hermite polynomials with order $i$ where $(i \in m,n)$, and the location of the peak is described by $(p,q)$, respectively.
The initial beam waist, Rayleigh length, and radius of curvature of the beam are defined as, $w_{c}$, $z_{R}=\pi w^{2}_{c}/\lambda$, and $R(z)=z+(z^{2}_{R}/z)$, respectively.
The beam spreads as it propagates through free space and obeys $w_{c}(z)=w_c\sqrt{1+(z/z_R)^2}$ where $z$ is the propagation distant.
Now the spatial effect of the control beam on the phase dependent probe absorption can be investigated by considering the shape of the control beam to be in $HG_{10}$ mode.
The transverse intensity distribution of the $HG_{10}$ mode about the $y$-axis is a two petal structure \cite{Milonni}.
Hence this two petal structure induces inhomogeneity in the probe transmission and creates spatial asymmetric transparency windows in the azimuthal plane as shown in Fig.(\ref{fig:Fig2})(b).
To extend the selective spatial transparency mechanism for arbitrary locations, we further use $HG_{11}$ mode control beam to illustrate the spatial modulation of the probe absorption.
Fig.(\ref{fig:Fig2})(c) exhibits four transparency windows located in the azimuthal plane at $(2n+1)\pi/4$ where $n:0\rightarrow 3$  due to the application of the $HG_{11}$ control field.
Further we show that the formation of asymmetric transparency window at a desired place can be decided by the peak position $(p,q)$ of $HG_{11}$ mode of the control field as depicted in Fig.(\ref{fig:Fig2})(d).
Therefore,  the results from Fig.(\ref{fig:Fig2})(b), (c) and (d) confirm that the spatial profile of the control beam and weak TMF play a decisive role in the creation of the asymmetric absorption pattern.
Also the probe absorption modulation can be controlled  by the intensity and the width of the control beam.
Hence such optical manipulation in the absorption can open exciting prospects for the creation of high contrast localized structured beams.

To delineate the effect of transverse profile of the control field  on the generation of structured beams, we study the Maxwell's equation under slowly varying envelope and paraxial wave approximations.
The equations in terms of probe Rabi frequencies $g_{\pm}$ are given by
\begin{align}
\label{probe}
 \frac{\partial g_{\pm}}{\partial z}
   &= \frac{\ii }{2{k_p}} \left( \frac{\partial^2 }{\partial x^2}
      + \frac{\partial^2 }{\partial y^2} \right) g_{\pm} + 2i{\pi}k_p{\chi}_{\pm}\,{g}_{\pm} \,,
\end{align}
where the first terms on the R.H.S. of Eq.~(\ref{probe}) account for diffraction of the probe beam, while the second terms on the R.H.S. leads to absorption and dispersion of the probe beam. The Fourier split-step operator method has been adopted to solve  the beam propagation equation numerically.
The initial spatial profile of the probe components are considered to be Laguerre-Gaussian with equal and opposite OAM and this can be expressed as:
\begin{align}
g_{\pm}(r,\phi,z=0)= g_{0}\left(\frac{r\sqrt{2}}{w_{p}}\right)^{\left|l\right|} e^{-\left(\frac{r^{2}}{w^{2}_{p}}\right)} L^{l}_{m}\left(\frac{2r^{2}}{w^{2}_{p}}\right) e^{\pm(il\phi)},\, 
\end{align}
\begin{figure}[h]
\includegraphics[width=8.5cm,height=7.5cm]{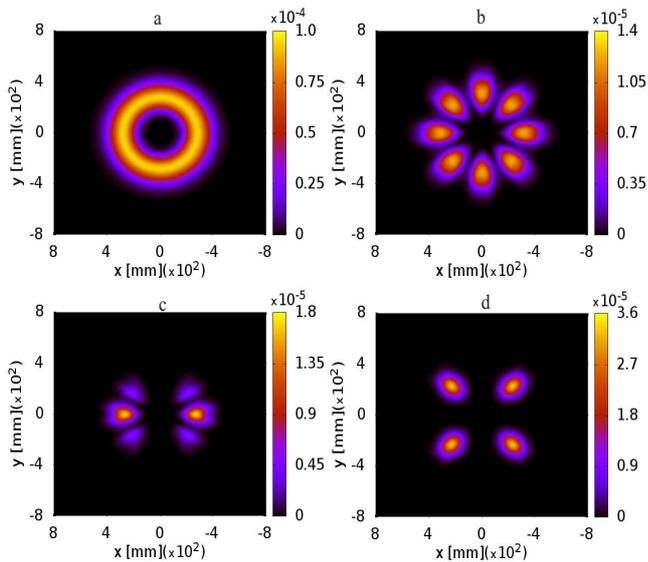}
\caption{ \label{fig:Fig3} (Color online) 
Transverse variation of probe intensity at different propagation distance. 
Panel (a) shows the input profile of the probe beam. 
Panel (b), (c) and (d) depict the creation of different structured beam patterns by control field induced asymmetric absorption. 
The beam parameters for the probe are $g_{0}=0.005\gamma$, $m=0$, $l=4$ and $w_{p}=20~\mu$m. 
The distance of propagation for the panel (b), (c) and (d) is $z=0.5~mm$.
Other parameters are same as in Fig.(\ref{fig:Fig2}).}
\end{figure}
where the radial distance from the axis of the beam is given by  $r=\sqrt{x^2+y^2}$ and the radial and azimuthal modes of Laguerre-Gaussian beam are described by $(m, l)$.
The output intensity can be evaluated using the expression $I_{out}= \left|g_{+}\right|^{2} + \left|g_{-}\right|^{2}$. 
The cross terms are absent from $I_{out}$ because of the orthogonality of the components of the probe field.
Figure~\ref{fig:Fig3}(a) shows the intensity distribution of the weak probe field against radial positions $x$ and $y$ at the entrance of the medium.
The azimuthal symmetry of Laguerre-Gaussian beam gives rise to the doughnut shape. 
Fig.(\ref{fig:Fig2})(a) suggests that the TMF initiates phase information modulation in the absorption spectrum that can be imprinted on the doughnut-shaped probe beam during its propagation.
However due to the presence of a huge medium absorption, the structure beam formation fails in the absence of the control field. 
It is clear from Fig.(\ref{fig:Fig3})(b) that the cw control field allows the mapping of the TMF induced phase information on to the doughnut-shaped probe beam which forms a 8 petalled structure.
We consider different modes of the Hermite-Gaussian control beam in order to demonstrate spatially asymmetric structure beam formation. 
Fig.(\ref{fig:Fig3})(c) shows how the $HG_{10}$ mode of the control beam encodes the phase information on to the probe profile around the two azimuthal positions, $0$ and $\pi$.
As a result, a 2 petalled spatial inhomogeneous structure is formed.
It is also evident from Fig.(\ref{fig:Fig3})(d) that the $HG_{11}$ control beam creates a diagonal structured beam consisting of 4 petals. 
Fig.(\ref{fig:Fig3})(c) and (d) illustrates that the peak transmission of the structured beam is found to be $18\%$ and $36\%$, respectively after a $0.5~$mm length of propagation, in the presence of a Hermite-Gaussian control field.
The enhancement of the generated structured beam features can be possible by suitably choosing the amplitude and the width of the Hermite-Gaussian control beam.
\begin{figure}[h]
\centering\includegraphics[width=6cm,height=5 cm]{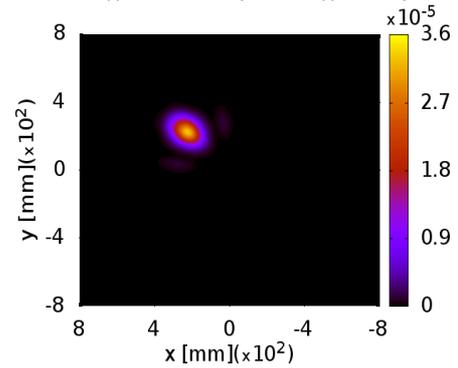}
\caption{\label{fig:Fig4} (Color online) 
Creation of a single spot using a shifted $HG_{11}$ control beam after propagation of $z=0.5~mm$. Other parameters are same as in Fig.(\ref{fig:Fig2}).}
\end{figure}
Hence the modes of the control beam behave like a phase selective tool in controlling and manipulating the spatial features of the structured probe beam. 
Finally we explore the possibility of selecting a single petalled structure of the probe beam at a desired location.
For this purpose, we have used a displaced $HG_{11}$ control beam mode.
The displaced control beam allows us to select a particular transparency window at an azimuthal plane and it creates a single petalled structure as shown in Fig.(\ref{fig:Fig4}).
Therefore, a spatially modulated control beam enables us to imprint TMF induced phase information at a desired location of the OAM carrying probe beam.
This leads to the formation of  asymmetric narrow features and high-contrast of structured beams, which have immense applications in imaging, biosciences, and in manipulation and trapping of particles \cite{Dunlop}.

In conclusion, we have carried out a theoretical investigation on the generation of structured beams at a desired transverse location in an ensemble of $^{87}$Rb atoms with inverted Y-level configuration.
A weak TMF and a suitable spatially dependent control beam have been used for the production of a spatial modulation of transparency that holds the key to  the formation of such sculpted light.   
This spatial absorption modulation enables us to imprint TMF induced phase information onto the probe beam at a specific azimuthal position which can be controlled by the spatial profile of the control field. 
Further the paraxial probe beam propagation equation unambiguously confirms that the control field manipulated selective phase information transfer can be efficiently cast onto the probe profile that leads to the generation of petalled structured beams. 
Thus this scheme can be utilized for the realization of optical tweezers \cite{Ashkin}, in high contrast imaging systems \cite{Geng}, and in micromachining \cite{Grier}.

\end{document}